\documentclass[twocolumn,prl,showpacs,superscriptaddress,amsmath,amssymb]{revtex4-1}

\usepackage{graphicx}
\usepackage{dcolumn}
\usepackage{bm}

\begin{document}


\title{Extended Phonon Collapse and the Origin of the Charge-Density-Wave in NbSe$_2$}


\author{F. Weber}
\email{frank.weber@kit.edu}
\affiliation{Materials Science Division, Argonne National Laboratory, Argonne, Illinois, 60439, USA}
\affiliation{Karlsruher Institut f\"ur Technologie, Institut f\"ur Festk\"orperphysik, P.O.B. 3640, D-76021 Karlsruhe, Germany}
\author{S. Rosenkranz}
\author{J.-P. Castellan}
\author{R. Osborn}
\affiliation{Materials Science Division, Argonne National Laboratory, Argonne, Illinois, 60439, USA}
\author{R. Hott}
\author{R. Heid}
\author{K.-P. Bohnen}
\affiliation{Karlsruher Institut f\"ur Technologie, Institut f\"ur Festk\"orperphysik, P.O.B. 3640, D-76021 Karlsruhe, Germany}
\author{T. Egami}
\affiliation{Department of Materials and Engineering, University of Tennessee, Knoxville, Tennessee, 37996, USA}
\author{A. H. Said}
\affiliation{Advanced Photon Source, Argonne National Laboratory, Argonne, Illinois, 60439, USA}
\author{D. Reznik}
\affiliation{Karlsruher Institut f\"ur Technologie, Institut f\"ur Festk\"orperphysik, P.O.B. 3640, D-76021 Karlsruhe, Germany}
\affiliation{Department of Physics, University of Colorado at Boulder, Boulder, Colorado, 80309, USA}


\date{\today}

\begin{abstract}
We report inelastic x-ray scattering measurements of the
temperature dependence of phonon dispersion in the prototypical
charge-density-wave (CDW) compound NbSe$_2$. Surprisingly,
acoustic phonons soften to zero frequency and become overdamped
over an extended region around the CDW wavevector. This
extended phonon collapse is dramatically different from the
sharp cusp in the phonon dispersion expected from Fermi surface
nesting. Instead, our experiments combined with ab initio
calculations, show that it is the wavevector dependence of the
electron-phonon coupling that drives the CDW formation in
NbSe$_2$ and determines its periodicity. This mechanism
explains the so far enigmatic behavior of CDW in NbSe$_2$ and
may provide a new approach to other strongly correlated systems
where electron-phonon coupling is important.
\end{abstract}


\maketitle

The origin of CDW order is a long-standing problem relevant to
a number of important issues in condensed matter physics, such
as the role of stripes in cuprate
superconductivity\cite{Kivelson03} and charge
fluctuations in the colossal magnetoresistive
manganites\cite{Dagotto05}. Static CDW order, i.e., a periodic
modulation of the electronic density, reflects an enhancement
of the dielectric response of the conduction electrons at the
CDW wavevector, $\bold{q}_{CDW}$, but it has long been known
that it is only stabilized by a coupling to the crystal
lattice\cite{Gruener88,Chan73}. Transitions into the CDW phase
on lowering the temperature are accompanied by a softening of
an acoustic phonon at $\bold{q}_{CDW}$ to zero frequency at
$T_{CDW}$ where it freezes into a static
distortion\cite{Hoesch09a} and evolves into the new periodic
(often incommensurate) superstructure. Chan and Heine derived
the criterion for a stable CDW phase with a modulation
wavevector $\bold{q}$ as\cite{Chan73}
\begin{equation}\label{equ1}
\frac{4\eta_q^2}{\hbar\omega_q}\ge\frac{1}{\chi_q}+(2U_q-V_q)
\end{equation}
where $\eta_q$ is the electron-phonon coupling associated with
a mode at a frequency of $\omega_q$, $\chi_q$ is the dielectric
response of the conduction electrons, and $U_q$ and $V_q$ are
their Coulomb and exchange interactions. Although both sides of
this inequality are essential in stabilizing the CDW order, the
common assumption is that the modulation wavevector,
$\bold{q}_{CDW}$, is determined by the right-hand side, i.e.,
by a singularity in the electronic dielectric function $\chi_q$
at $\bold{q}_{CDW} = 2k_F$, where $k_F$ is the Fermi
wavevector. In the case of NbSe$_2$, it was proposed that such
a singularity resulted either from direct Fermi surface nesting
at $\bold{q}_{CDW}$ \cite{Gruener88,Wilson77} or from the
presence of saddle points near the Fermi surface connected by
$\bold{q}_{CDW}$ \cite{Rice75}. However, this has been
challenged by a recent density-functional-theory (DFT)
calculation that correctly predicted a CDW instability, but
without singularities in $\chi_q$ \cite{Johannes06,Calandra09}.
Furthermore, several angle-resolved photoemission spectroscopy
(ARPES) experiments have found no clear evidence of an energy
gap at the nesting wavevectors opening at $T_{CDW}$
\cite{Shen08,Borisenko09,Rossnagel01}. As a consequence, the
elegant and intellectually compelling picture of CDW formation
driven by Fermi surface nesting has been called into
question\cite{Johannes06}. The alternative possibility is that
the CDW transition in NbSe$_2$ may instead be driven by the
left-hand side of equation 1, i.e., the wavevector dependence
of the electron-phonon coupling $\eta_q$. A direct test of this
conjecture by phonon spectroscopy is the subject of this
letter.

We have investigated discrepancies with the Fermi
surface-nesting scenario by measuring the energies and
linewidths of phonon excitations in NbSe$_2$ as a function of
temperature, comparing both to ab initio calculations. In particular the phonon linewidth, which is
proportional to the product of the electron-phonon coupling
$\eta_q$ and the electronic joint density-of-states (JDOS),
which in turn is related to $\chi_q$, provides detailed
information related to the Fermi surface. Our results
demonstrate that the soft phonon physics in NbSe$_2$ does
indeed deviate strongly from the conventional picture based on
nesting\cite{Gruener88,Hoesch09a} and that the electron-phonon
coupling is primarily responsible for determining the wave
vector of CDW order.

\begin{figure}
\begin{center}
\includegraphics[width=0.8\linewidth]{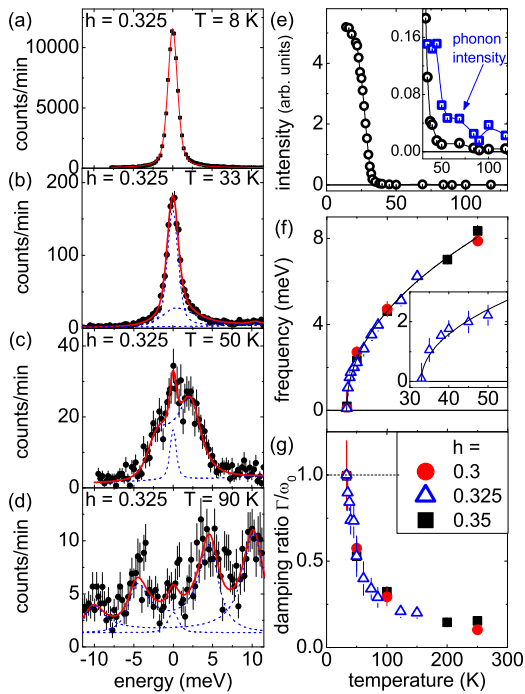}
\caption{\label{fig_1} (color online) Temperature dependence of the soft phonon mode and the charge-density-wave superlattice peak near $\bold{q}_{CDW} = (0.329, 0, 0)$. \textit{(a)-(d)} Energy scans at $\bold{q} = (3-h, 0, 0)$, $h = 0.325$, for temperatures $8\,\rm{K}\le\rm{T}\le90\,\rm{K}$. Solid (red) lines are fits consisting of  damped harmonic oscillators (inelastic) and a pseudo-voigt function (elastic) (blue dashed lines). \textit{(e)} Intensity of the charge-density-wave superlattice peak for $T\,\le\,120\,\rm{K}$. The inset shows the phonon and superlattice peak intensities just above $T_c$. \textit{(f)(g)} Phonon frequency $\omega_q$ and critical damping ratio $\Gamma/\omega_0$ of the soft phonon mode, respectively, at $\bold{q} = (h, 0, 0)$ with $h = 0.3$ (circle), $0.325$ (triangle) and $0.35$ (square). The solid line in \textit{(f)} is a power law fit of the form $((T-T_c)/T_c)^{\delta}$ yielding $\delta = 0.48 \pm 0.02$.}
\end{center}
\end{figure}

Earlier inelastic neutron scattering investigations of the
phonon dispersion in NbSe$_2$ were limited by weak scattering
intensity due to the small sample size and by relatively poor
wavevector resolution\cite{Moncton75,Ayache92}. To overcome
these constraints, we utilized high-resolution inelastic X-ray
scattering (IXS), which allowed us to obtain measurements of
the entire dispersion of the soft mode branch in a wide
temperature range. We used a high-quality single crystal sample
of about $50\,\rm{mg}$ ($2 \times 2 \times 0.05\,\rm{mm}^3$)
with a $T_{CDW}$ of $33\,\rm{K}$ determined from the
temperature dependence of the superlattice reflections (Figure
\ref{fig_1}) in agreement with previous
results\cite{Moncton75}. All IXS experiments were carried out
on the XOR 30-ID HERIX beamline at the Advanced Photon Source,
Argonne National Laboratory. Data were fitted with damped
harmonic oscillator functions convoluted with the experimental
resolution. For more details on the instrumental setup and data
analysis see the supplementary material. Here, we focus on the
longitudinal acoustic phonon branch dispersing in the
crystallographic $(100)$ direction and crossing the CDW wave
vector $\bold{q}_{CDW} = (0.329, 0, 0)$\cite{Moncton75}.
Figure \ref{fig_1} shows the temperature dependence of a soft phonon mode at $\bold{q}_{hkl} = (0.325, 0, 0)$, close to the CDW wavevector, $\bold{q}_{CDW} = (0.329, 0, 0)$. At $T = 90\,\rm{K}$ (Fig. \ref{fig_1}d), the soft phonon at an energy of $\omega_q = 4.5\,\rm{meV}$ has nearly equal intensity to the second phonon branch at $10\,\rm{meV}$. Upon cooling, the intensity of the upper branch is suppressed due to the Bose factor, whereas the intensity of the soft phonon is enhanced by a factor $1/\omega_q$ in the cross section as its frequency, $\omega_q$, is reduced. At $T = T_{CDW}$, the elastic superstructure peak of the CDW phase dominates the spectrum (Fig. \ref{fig_1}b), but we can still distinguish the critically damped phonon as a broad peak beneath the narrow elastic CDW peak. Well inside the CDW phase, the elastic superlattice reflection was too strong for any inelastic scattering to be observed at $\bold{q}_{CDW}$ and these low temperature data (Fig. \ref{fig_1}a) were used to determine the shape of the elastic scattering.

Figure \ref{fig_1}e shows that the integrated intensity of the CDW superlattice peak measured at $h = 0.325$, which is within the momentum-resolution of $\bold{q}_{CDW}$, increases rapidly below $T_{CDW} = 33\,\rm{K}$, in good agreement with previous neutron diffraction data taken on crystals from the same growth batch\cite{Moncton75}. Above $T_{CDW}$, the elastic intensity due to diffuse scattering from the sample is very small, which implies that our sample had very little structural disorder. It stays low until very close to $T_{CDW}$ (see inset to Fig. \ref{fig_1}e), where a weak elastic "central" peak consistent with low energy critical fluctuations appears.
\begin{figure}
\begin{center}
\includegraphics[width=0.7\linewidth]{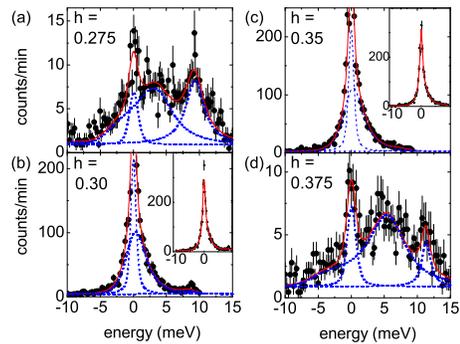}
\caption{\label{fig_2} (color online) Wavevector dependence of the soft-phonon at $T = 33\,\rm{K}$. Energy scans at $\bold{Q} = (3-h, 0, 1)$, $h = 0.275 - 0.375$. Solid (red) lines represent the total fit result consisting of a damped harmonic oscillator functions (inelastic) and a pseudo-voigt function (elastic) (blue dashed lines).}
\end{center}
\end{figure}
The phonon frequency at $\bold{q}_{CDW}$ softens on cooling (Fig. \ref{fig_1}f) following a power law $\omega_q(T) = ((T-T_c)/T_c)^\delta$ with $\delta = 0.48 \pm 0.02$, the value predicted by mean-field theory\cite{Gruener88}. As the phonon softens, the damping increases and the phonon becomes critically damped, i.e. $\Gamma/\omega_0 = 1$, at $T_{CDW}$ (Fig. \ref{fig_1}f).
\begin{figure}
\includegraphics[width=0.5\linewidth]{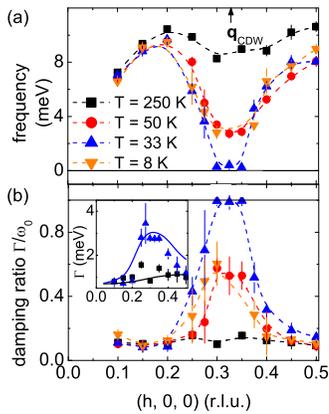}
\caption{\label{fig_3} (color online) Experimentally obtained dispersion and damping ratio of the soft phonon branch in NbSe$_2$ at four temperatures $8\,\rm{K}\le\rm{T}\le250\,\rm{K}$. Plotted are \textit{(a)} the frequency of the damped harmonic oscillator (DHO) $\omega_q = \sqrt{\omega_0^2-\Gamma^2}$ and \textit{(b)} the damping ratio $\Gamma/\omega_0$. Lines are guides to the eye. Note that phonons at $h = 0.325$, $0.35$ and $T = 8\,\rm{K}$ were not detectable due to strong elastic intensities. The inset in \textit{(b)} shows the experimentally observed damping $\Gamma$ of the damped harmonic oscillator (symbols) and scaled DFPT calculations (see Fig. \ref{fig_4}) of $2\gamma$ (lines, offset $0.7\,\rm{meV}$) with $\sigma = 0.1\,\rm{eV}$ (black) and $1\,\rm{eV}$ (red).}
\end{figure}
Remarkably, we observe the same power law behavior not only at
$\bold{q}_{CDW}$, but also at $h = 0.3$ and $0.35$ which are
far outside the experimental resolution from $\bold{q}_{CDW}$
and where the elastic peak is an order of magnitude weaker
relative to the phonon intensity. Moreover, the phonon
frequencies at these wavevectors also become indistinguishable
from zero at $T_{CDW}$ (Fig. \ref{fig_2}b,c) the same as at
$\bold{q}_{CDW}$. This means that the phonons are critically
damped over a large range of momentum transfer from $h=0.3$ to
$h=0.35$. Going further away from $\bold{q}_{CDW}$ with the
same step size, $\Delta h = 0.025\,\rm{r.l.u.}$, the soft
phonon branch is well separated from zero energy (Fig.
\ref{fig_2}a,d). Figure \ref{fig_3} shows the full dispersion
and damping ratio of the soft-mode phonon branch. A broad
dispersion anomaly is already evident at $250\,\rm{K}$ in
agreement with previous neutron scattering measurements
performed only at $300\,\rm{K}$\cite{Moncton75}. This anomaly
deepens considerably upon cooling to $50\,\rm{K}$, where we
also observe a strong increase in the damping. Finally, upon
cooling to $T_{CDW}$, the frequencies reach zero and the
phonons become critically damped over an extended range of
wavevectors. At $T = 8\,\rm{K}$, well below $T_{CDW}$, we find
hardened frequencies and reduced damping, similar to the ones
observed at $T = 50\,\rm{K}$. However, the soft mode was not
resolvable at $h = 0.325$ and $0.35$ due to strong elastic
scattering from the CDW peak (Fig. \ref{fig_1}a). At these
temperatures the Bose and $1/\omega$ factors suppress the
phonon intensity and the measurements become increasingly
difficult.

\begin{figure}
  \begin{center}
    \includegraphics[width=0.6\linewidth]{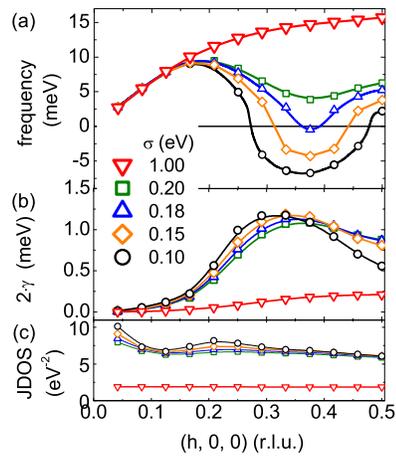}
    \caption{\label{fig_4} (color online) Ab initio calculation for the soft phonon mode along the crystallographic $(100)$ direction in NbSe$_2$. Shown are the calculated \textit{(a)} dispersion, \textit{(b)} $2\gamma$ (the contribution of the electron-phonon interaction to the phonon linewidth (FWHM)), and \textit{(c)} the electronic joined density of states (JDOS). Calculations were done with five different Gaussian broadenings $0.1\,\rm{eV} \le \sigma \le 1.0\,\rm{eV}$ (see text). Note that calculated imaginary frequencies are shown as negative roots of the square phonon frequencies. Lines are guides to the eye.}
  \end{center}
\end{figure}
The $\bold{q}$-dependence of the phonon softening shown in
Figure \ref{fig_3} is in marked contrast to the sharp,
cusp-like dips that normally characterize Kohn anomalies at
$2\bold{k}_F$ due to Fermi surface
nesting\cite{Hoesch09a,Bohnen04}. For example in the CDW
compound ZrTe$_3$, where the importance of Fermi surface
nesting has been clearly established\cite{Hoesch09b}, phonon
frequencies are only renormalized over a $\bold{q}$-range of
$0.085\,$\AA$^{-1}$ and soften to zero at a single wavevector
using a step size of $0.04\,$\AA$^{-1}$\cite{Hoesch09a}. In
contrast, in NbSe$_2$, we find that the phonon renormalization
extends over $0.36\,$\AA$^{-1}$, or over half the Brillouin
zone, and the critically damped region extends over
$0.09\,$\AA$^{-1}$, about twice the experimental resolution.
This behaviour clearly rules out a singularity in the
electronic response in NbSe$_2$ and suggests that the CDW is
determined by the wavevector dependence of the electron-phonon
coupling $\eta_q$, as proposed by Johannes et
al.\cite{Johannes06}. A broadened or even flat-topped
susceptibility due to imperfect nesting caused, e.g., by the
c-axis dispersion of the electron bands, could also lead to a
renormalization of the phonon dispersion over a larger range of
wavevectors, but it is unlikely that it spans over half of the
Brillouin zone.

In order to elucidate the microscopic mechanism behind the CDW
phase transition in NbSe$_2$, we compare our experimental
results to detailed phonon calculations
based on \textit{density functional perturbation theory} (DFPT)
performed with the crystal structure at $T > T_{CDW}$. This is
a zero temperature technique, in which structural instabilities
show up as imaginary phonon frequencies (see supplementary
material). Because of the finite momentum mesh used in the DFPT
calculations a numerical smearing, $\sigma$, of the electronic
bands is necessary to compare the calculations with experiment.
The effect of $\sigma$ is analogous to a thermal smearing of
the electronic structure, so it has been used in previous work
to simulate the effect of
temperature\cite{Bohnen04,Pintschovius08}. However, a
na\"{\i}ve interpretation, i.e., $\sigma =
2.12*k_BT$\cite{Bohnen04}, results in temperatures at least one
order of magnitude too large with respect to experimental
observations\cite{Pintschovius08}. We believe that the
calculated temperatures disagree quantitatively with experiment
because the calculations ignore the thermal disorder of the
lattice, which can have profound effects on the phonon
excitations\cite{Delaire08}. Elucidating these issues is an
important direction for future research beyond the scope of the
present work. We merely note that for NbSe$_2$ a comparison
between theory and experiment indicates that values of
$0.1\,\rm{eV} \le \sigma \le 1\,\rm{eV}$ produce results that
are consistent with a temperature range of $30\,\rm{K} \le T
\le 300\,\rm{K}$.

Figure \ref{fig_4} summarizes the calculations, showing the
calculated soft-phonon dispersion, linewidth, and electronic
joint density of states (JDOS). Imaginary phonon-frequencies
are represented in Fig. \ref{fig_4}a through the negative roots
of the absolute value, e.g., as 'negative' phonon frequencies.
These occur in the calculated longitudinal acoustic phonon
branch for $\sigma \ge 0.18\,\rm{eV}$ over an extended range of
wavevectors (Fig. \ref{fig_4}a) in agreement with previous
studies\cite{Calandra09}, and in a qualitative agreement with
the observed breakdown of the phonon dispersion. Similarly, the
contribution to the phonon linewidth from the electron-phonon
interaction, $2\gamma$ (Fig. \ref{fig_4}b), shows a strong
enhancement over the same extended range of wavevectors.
$2\gamma$ is proportional to the product of the electron-phonon
coupling and the electronic joint density of states (JDOS).
Since the latter shows negligible wavevector dependence (Fig.
\ref{fig_4}c), the enhancement of the phonon linewidth observed
in both experiment and theory is entirely due to a strong
wavevector dependence of the electron-phonon coupling.
Moreover, this range of strongly enhanced electron-phonon
coupling is identical to the range over which the phonon
softens. In contrast, the real part of the susceptibility only
shows a much broader and only very shallow
peak\cite{Johannes06}. This leads us to conclude that the
observed $\textbf{q}$-dependence of the phonon self-energy is
entirely due to the electron-phonon coupling and that the CDW
wavevector in NbSe$_2$ is indeed determined by the left-hand
side of equation \ref{equ1}.

Since our measurements demonstrate that Fermi surface nesting
does not play a role in the CDW phase transition in NbSe$_2$,
the electronic states serve only to provide an elevated
dielectric response, with the modulation wavevector entirely
determined by the coupling between electronic and vibrational
dynamics. Previous studies of chromium\cite{Lamago10} and
ruthenium\cite{Heid00} have shown that matrix elements can
indeed be sharply wavevector dependent and also produce dips in
the phonon dispersions, although in these compounds phonons do
not soften to zero energy.  Our work provides direct evidence
that the same effect can drive the structural instability in a
CDW compound. This result naturally explains why electronic
probes do not find strong nesting or a clear gap at
$\bold{q}_{CDW}$ in NbSe$_2$. Our results have implications for many other strongly correlated systems.
In particular, CDW correlations in the form of stripes and/or
checkerboard patterns have been linked to the emergence of
unusual states and physical properties, such as colossal
magnetoresistance in the manganites\cite{Cox07} and the
pseudogap state in the cuprates\cite{Damascelli03}. Indeed, the
observation of phonon anomalies in manganites at the wavevector
of the checkerboard-type order\cite{Weber09} and anomalies
observed in La$_{2-x}$Sr$_x$CuO$_4$ at the stripe ordering
wavevector\cite{Reznik06} demonstrate that strong
electron-phonon coupling could be important in these materials
as well.

In conclusion, we reported inelastic x-ray measurements of the
temperature dependence of a longitudinal acoustic phonon in
NbSe$_2$ involving the CDW soft mode. We observe an extended
region in $\textbf{q}$ with overdamped phonons at the CDW
transition temperature. A detailed comparison to lattice
dynamical calculations via DFPT shows that in NbSe$_2$, the
periodicity, i.e. $\bold{q}_{CDW}$, of the CDW ordered state is
determined entirely by the wavevector dependence of the
electron-phonon coupling. This is in stark contrast to the
standard view that a divergent electronic response defines
$\bold{q}_{CDW}$ and is evidence that a wavevector dependent
electron-phonon coupling can drive a structural phase
transition.

\begin{acknowledgments}
We acknowledge valuable discussions with Igor Mazin, Jasper van
Wezel, and Mike Norman. We thank John M. Tranquada for
supplying us with a single crystal of NbSe$_2$.  Work at
Argonne was supported by U.S. Department of Energy, Office of
Science, Office of Basic Energy Sciences, under contract No.
DE-AC02-06CH11357.
\end{acknowledgments}


\end{document}